# Synthetic exposure: a simplified and accurate acquisition scheme for multiple exposure speckle imaging of blood flow

[1]Chammas M, *,[1]Pain, F

[1]Université Paris-Saclay, Institut d'Optique Graduate School, CNRS, Laboratoire Charles Fabry, 91127, Palaiseau, France.

* corresponding author: frederic.pain@universite-paris-saclay.fr

**Abstract:** Speckle contrast imaging is an established technique to obtain relative blood maps over wide field of views. Currently, its most accurate implementation relies on the acquisition of raw speckle images at different exposure times but requires modulation of a laser pulse in duration and intensity and precise synchronization with camera. This complex instrumentation has limited the use of multiple exposure speckle imaging. We evaluate here a simplified approach based on synthetic exposure images created from the sum of successive frames acquired with a 1 ms exposure time. Both methods have been applied to evaluate controlled flows in micro-channels. The contribution of noises to the speckle contrast have been quantified and compared. Dark, readout and shot noise contributions to the total contrast remains constant for modulated exposure, while all these contributions decrease with increasing exposure time for synthetic exposure. The relative contribution of noises to speckle contrast depends on the level of illumination, the exposure time and the flows that are imaged. Guidelines for accurate flow measurements in the synthetic exposure acquisition scheme are provided. The synthetic exposure method is simple to implement and should facilitate the translation of multiple exposure speckle imaging to clinical set-ups.

## 1. Introduction

Speckle contrast imaging is an efficient tool to monitor *in vivo* relative blood flow changes and tissue perfusion. It has been used in several medical fields such as rheumatology, dermatology, ophthalmology, and neurology [1]. Multiple exposure speckle imaging (MESI) has proven to be more accurate than simple exposure laser-speckle imaging for flow quantification due to its ability to differentiate the contribution of static scatters (i.e. bones, cartilages) from that of the moving scatterers (i.e. red blood cells) to the speckle patterns. In addition, MESI retrieves accurate flows over a much broader linear range and is especially more accurate for large flow variations[2,3]. Yet most laboratory and commercial laser speckle contrast imagers still rely on the simple exposure approach as its practical implementation is much simpler [4,5]. In the MESI system developed previously in our group[6], the acquisition of multiple exposure relies on a fixed camera exposure associated to laser pulses varying in duration and intensity as proposed previously[7]. This implies the use of an acousto-optic modulator (AOM) associated to an arbitrary waveform generator (AWG) and the precise synchronization of the camera to the laser pulses. We investigate here a much simpler and cost-effective instrumental approach. Synthetic exposures images for different exposure times are obtained by the summation of images acquired at a short exposure time (e.g. a 10 ms synthetic exposure image is obtained by the summation of 10 images obtained at 1 ms). This approach has been proposed previously using high sensitivity, low noise SPAD detectors but at the expense of a strongly limited spatial resolution[8]. Further studies have used synthetic exposure with fast CMOS cameras but no systematic comparison to the established laser modulation method was performed as a validation[9,10]. On one hand, the camera noises (dark, read-out and shot noise) in each image add with the increasing exposure time in the synthetic exposure mode. On the other hand, the synthetic exposure approach is likely to be much less sensitive to the intensity calibration bias observed in the modulated laser approach.

Indeed, in this approach, not only the duration of each pulse is modulated, but also their amplitude to keep a similar light level at the detector for all exposure times. This requires a priori calibration, which introduces ultimately uncertainty on the blood flow evaluation. The present study was designed to compare the synthetic exposure approach to the laser modulation approach in typical conditions associated with blood flow imaging in mice brain [11].

## 2. Results

In dynamic speckle imaging the local contrast is estimated as the ratio of the local standard deviation divided by local mean of pixels intensities (see Eq (1) in section 4.2). The underlying concept is that for fast flows, the speckle patterns decorrelates quickly resulting in a low local contrast, while for static areas, the speckle patterns remain identical, and the local contrast is high. For speckle contrast imaging of a given flow under a fixed illumination, variances in the pixel intensities arise not only from the dynamic speckle patterns fluctuation related to the flow but also from noises (dark, readout, shot noises). In synthetic exposure, the successive addition of frames results in adding the different variances contribution for each pixel. To correct for the contributions not related to flow (i.e. noises) the contributions of noises should be estimated thoroughly.

### 2.1. Experimental characterization of dark and readout noise for a sCMOS sensor

Figure 1 shows the experimental characterization of the dark and readout noises of the Orca Flash 2.8 camera operated in simple exposure mode (e.g., each exposure time corresponds to the acquisition of a single independent image at this exposure). As can be seen, the dark and readout noises are low and homogenous for all the pixels (Figure 1 A, B). In addition, the dark and readout noises are spatially homogeneous (Figure 1C) and temporally invariant (Figure 1D) making the subtraction of this contribution straightforward for this specific camera.

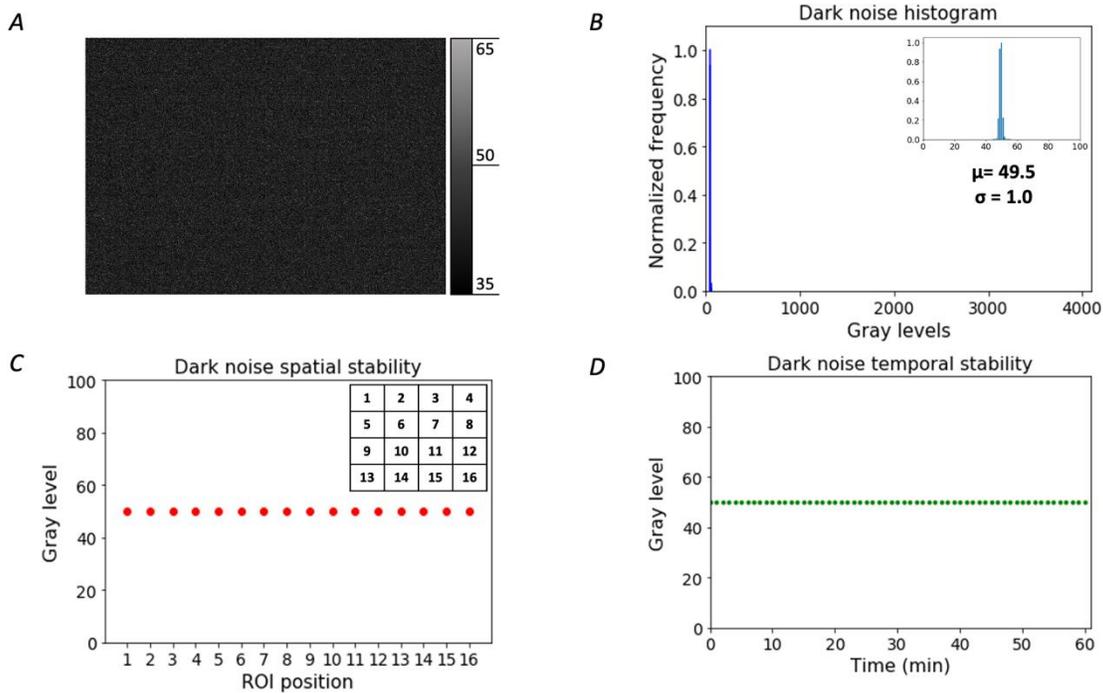

Figure 1 Camera dark and readout noise homogeneity and stability A) Representative dark image B) Corresponding histogram of grey levels. Insert is a zoom on non-zero histogram levels C) Spatial uniformity of noise, the sensor was divided in 16 regions of identical dimensions numbered from 1 to 16 D) Temporal stability of noise evaluated as the mean grey level of the whole sensor.

## 2.2. Estimation of signal standard deviation contributions in modulated laser and synthetic exposure modes

The standard deviation of the signal intensities is at the core of the speckle contrast calculation. The direct subtraction of a noise image to the signal images allows correction of the pixel intensities but subtraction leads to an added variance. Therefore, the variance due to dark, readout and shot noises should be evaluated and removed from the raw variance to obtain an accurate estimation of the speckle contrast related to the velocity of the moving red blood cells. Figure 2 A, B show the experimental evaluation of the raw, dark and readout noise, and shot noise standard deviation as a function of the exposure time in the modulated laser mode (mode 1) and in the synthetic exposure mode (mode 2). The standard deviation of dark, readout and shot noise remain constant for all exposures for the modulated laser mode whereas standard deviation adds up with the number of frames in the synthetic exposure mode. In mode 1 the amount of detected light is maintained constant for all exposure times by adjusting the laser intensity with the duration of the pulses. Therefore, the shot noise standard deviation remains constant. Contrarily, in mode 2, the computed detected signal adds with the number of frames resulting in a much higher dynamic for the synthetic image than what is allowed by the limited depth well of the pixels. In other words, if a pixel is already close to intensity saturation for a single image at 1 ms exposure time, adding 10 frames will not saturate the synthetic 10 ms image like a real single 10 ms exposure would. Regarding the dark and readout, as only one image is read for each exposure times in mode 1, the readout and dark are independent of the exposure times, provided that no varying ambient light is recorded. In mode 2, individual frames are added to create long exposure times, so the dark and readout standard deviations increases with the number of frames used to compute the speckle image for each exposure time.

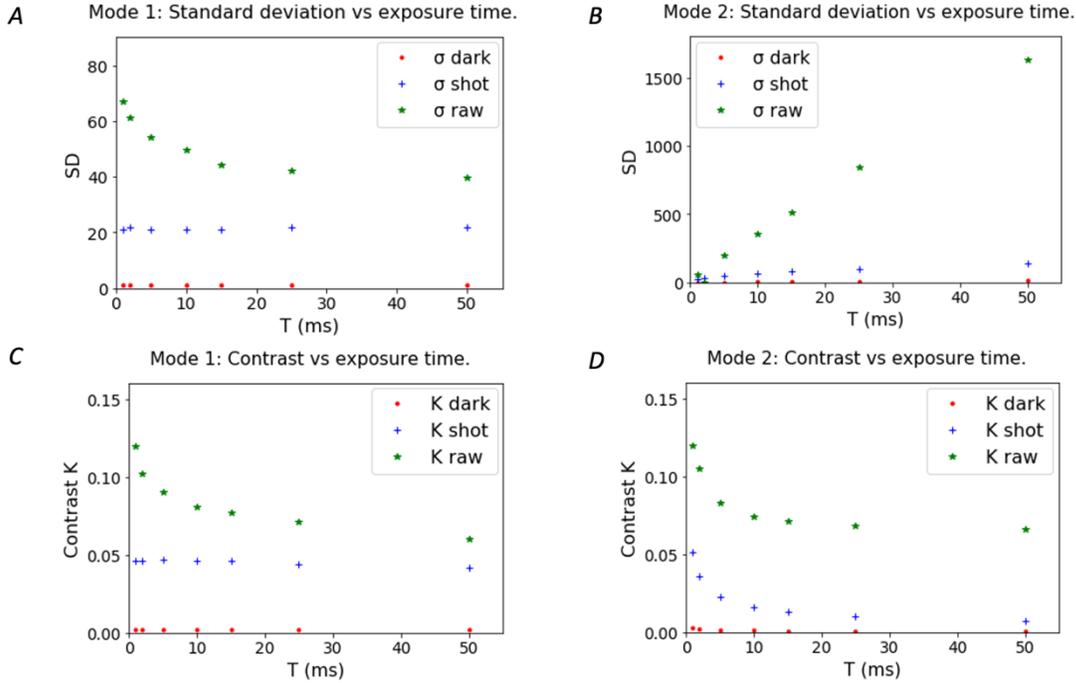

Figure 2. A) Raw, dark and readout, and shot noise standard deviations in the modulated laser mode as a function of exposure time. B) Same as A) for the synthetic exposure mode C) Raw, dark and readout and shot noise speckle contrast for the modulated laser mode D) Same as C) for the synthetic exposure mode.

## 2.3. Contributions of the noises to the speckle contrast in modulated laser and synthetic exposure modes

The metric of interest for flow evaluation in speckle imaging is the local speckle contrast. Figure 2 C, D shows the raw speckle contrast as well as the contribution of the dark and shot contrast for both acquisition modes as a function of exposure time (see section 4.4 for calculations methods). As expected, the contributions of the shot and dark remain constant throughout the whole range of exposure times for the modulated laser approach. On the contrary, for synthetic exposures, the contributions to the speckle contrast of the shot and the dark noise decrease with the exposure (i.e., the number of added frames). Indeed, the standard deviation for both contributions increases as the square root of the number of frames, whereas the signal increases linearly with the number of frames leading to a theoretical decrease of the contribution of each noise to the speckle contrast as $\frac{1}{\sqrt{N.I_{one\ frame}}}$. Where N is the number of frames and I one frame is the intensity level of one frame. Consequently, the contributions to the contrast of the shot and dark noise are significantly lower in mode 2 for all exposures except the shortest ones.

## 2.4. Relative contribution of the shot noise to the contrast as a function of illumination intensity

The relative contribution of the shot noise to the raw speckle contrast depends on the detected signal intensity. In mode 2, the shot noise contribution is determined by the level of detected signal at the lowest exposure time and the number of frames that are added to compute the speckle patterns for each exposure time. Figure 3 shows the influence of the noise correction for different levels of illumination of the sample, ranging from optimal use of the camera dynamic range (quasi saturation of the camera well depth) to sub-optimal level of signal (less than half of the pixels depth well are used). For optimal use of the camera dynamic, the contribution of the shot noise is almost negligible even for the shortest exposure time (Figure 3 A, illumination at 223 mW/cm$^2$, T = 1 ms). When the signal level is sub-optimal the relative contribution of the shot noise is significantly higher. This is likely to occur when speckle imaging is carried out over large field of view as the laser power is distributed over a large surface (figure 3 A, illumination at 20 mW/cm$^2$, T = 1 ms). Figure 3B, shows that the relative contribution of shot noise to the contrast increases also when short exposure times are required to catch the contrast variations due to large flows. As a summary, the insert in figure 3B shows that the contribution of shot noise to the speckle contrast rises to 15% when speckle patterns are acquired under low illumination and for short exposure time. On the contrary, this contribution is limited to less than 2% for all exposure times when optimal illumination is obtained.

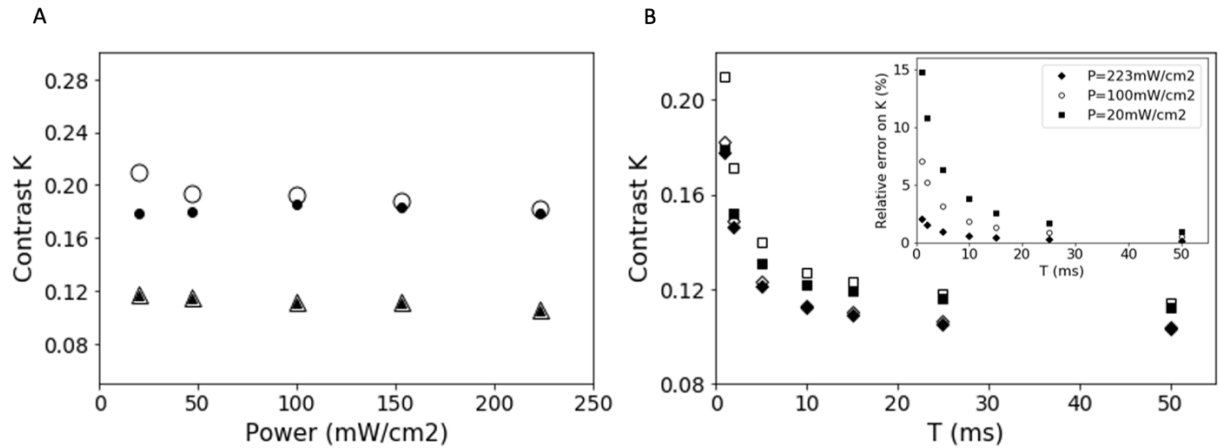

Figure 3. Effect of illumination level and exposure time on the shot noise contribution to the speckle contrast in synthetic exposition mode A) Effect of the detected light intensity; empty dots (raw contrast) and filled dots (noise corrected contrast) correspond to an exposure time of 1 ms ; empty triangles (raw contrast) and filled triangles (noise corrected contrast) correspond to an exposure time of 25 ms B) Effect of the exposure time; empty squares (raw contrast) and filled squares (noise corrected contrast) correspond to an illumination of 2 mW/cm$^2$; empty diamonds (raw contrast) and filled diamonds (noise corrected contrast) correspond to an illumination of 223mW/cm$^2$. Insert: Relative error due to the shot noise speckle contrast as a function of exposure time for different illumination levels; squares, dots and diamonds correspond respectively to illuminations at 20, 100 and 223mW/cm$^2$.

## 2.5. Relative contribution of shot noise to the contrast as a function of the measured flow

The relative contribution of shot noise to the speckle contrast depends also on the speckle contrast related to the flow that is imaged. In dynamic speckle imaging, high flows result in lower contrast whereas low flows lead to higher contrasts values. Consequently, for the same level of illumination, the relative contribution of the shot noise to the speckle contrast varies with the flow being imaged. Figure 4 shows that the relative contribution is much higher for a flow of 8µl/min compared to a flow 2µl/min for the whole range of exposure times considered, while as shown previously, the shot noise contribution decreases with the exposure time for all flows. Nevertheless, as data have been acquired under optimal illumination, it can be observed that the error remains below 2%

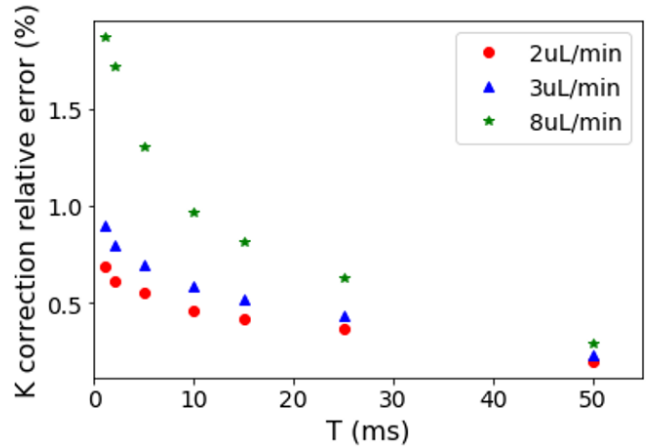

Figure 4. Relative error on contrast calculation due to noises contribution as a function of exposure time for different flows of intralipid 2% in a 300 μm diameter channel. Data have been acquired in synthetic exposure, with frames of 1 ms exposure time, and under optimal illumination using the full dynamic of the camera sensor.

## 2.6. Comparison of flow quantification from data acquired using modulated laser or synthetic acquisition

Data were acquired sequentially in both modes for controlled flows between 1 and 7 μl/min of 2 %-intralipid in a 300 μm microchannel. Representative speckle contrast images for 6 different exposure times are shown for acquisition mode 1 and mode 2 on figure 5A and figure 5B respectively. Figure 5C shows representative data corresponding to the speckle contrasts for a 3 μl/min flow. Speckle contrasts derived from both acquisition modes show similar values for the whole range of exposure times. The correlation times for different flows are derived from a fit of the corresponding data to equation (2) (see section 4.1). Figure 5D shows the relative correlation time obtained from modulated laser acquisition as a function of those obtained from synthetic exposure. There is a strong linear relationship between the data obtained from both acquisition modes, with a slope close to unity, showing the ability of synthetic exposure to measure accurately relative flow changes in a large range of flows.

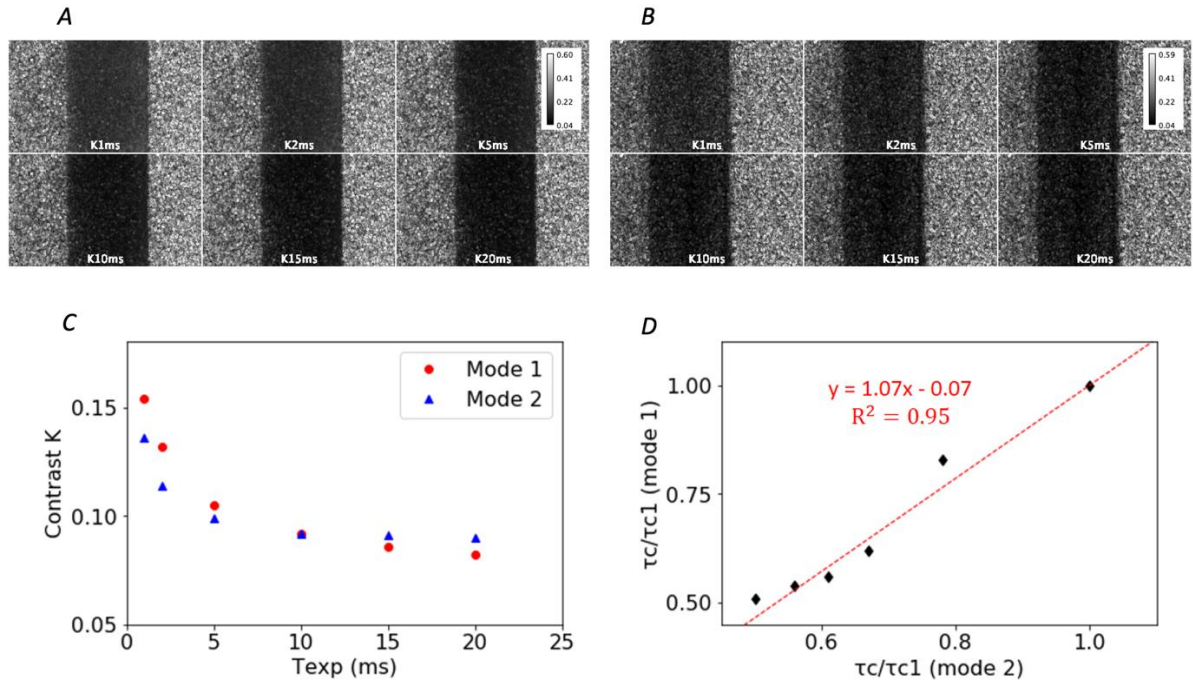

Figure 5. Representative comparison between mode 1 (Laser Modulated Method) and mode 2 (Synthetic Exposure) **A)** Speckle contrast images of intralipid 2% flowing at 3μl/min in a 300 μm diameter channel for 6 exposure times (mode 1) **B)** Same as (A) for mode 2 **C)** Speckle contrast versus exposure time **D)** normalized correlation times for mode 1 as a function of those of mode 2.

## 3. Discussion

### 3.1. Camera noise contributions to the speckle contrast

In the present study, we use a scientific CMOS camera (sCMOS), which was first issued by Hamamatsu in 2012. Its modest specifications compared to more recent sCMOS cameras do not hinder accurate MESI acquisitions in the synthetic exposure mode. Obviously, higher spectral efficiency, and lower dark and readout noise would further decrease the contribution of noise to the speckle contrast, yet the main noise contribution to the speckle contrast arises from the shot noise, which is inherently independent of the camera. However, specifications of common CMOS cameras, even from established manufacturers should be considered cautiously for speckle imaging in the synthetic exposure mode. Indeed, in CMOS cameras, the charge to voltage conversion is independent for each pixel and each column has its own amplifier and analog-to-digital converter, in addition to possible pixel-to-pixel variation in quantum efficiency. The "raw" output from the CMOS image sensor includes pixel-to-pixel variability in the read noise, electronic gain, offset and dark current, which are corrected by the manufacturer in high performance sCMOS cameras[12] but not in regular CMOS cameras. Consequently, standard CMOS usually have moderate to strong inhomogeneous spatial noise (quantified as Dark Signal Non-Uniformity, DSNU), spatial patterns of noise arising from the architecture of the electronics, which should be considered cautiously if systematic noise correction is carried out. Although lasers speckle imaging set up have been implemented with moderate performance cameras like webcams, it is unlikely that such approach would allow accurate multiple exposure imaging in the synthetic exposure mode[13] as the successive frames summation would make the uncorrected noise pattern more visible and ultimately bias locally the speckle contrast calculation.

### 3.2. Illumination level and uniformity

A general guideline for laser speckle imaging is that the illumination of the imaged tissues should be set to maximize the use of the camera dynamics and to minimize the shot noise contribution. The shot noise contribution to the contrast is maximal when the illumination level is sub-optimal (Figure 3). In the synthetic exposure mode this can be true for low exposure times while longer exposure times benefit from the summation of multiple frames. Similarly, in the study by Valdes et al [14], shot noise contribution overwhelms the flow-related speckle contrast for tissues areas with a low signal to noise ratio. The contribution of shot noise under low illumination could be problematic in the case of clinical application of speckle contrast imaging as practical and safety constraints impose the use of low power laser beams expanded over wide field of view thus limiting the amount of signal [15]. In these conditions, the synthetic exposure mode could be particularly interesting compared to the modulated laser mode. It would allow to strongly decrease the shot noise contribution to the speckle contrast for long exposure times requested to probe small flow changes. The inter pixel variability in sensitivity and electronic conversion chain is quantified as the photon response non uniformity (PRNU) for CMOS cameras. In sCMOS cameras, it is carefully corrected[12] similarly to DSNU avoiding unwanted spatial patterns that could bias the speckle images in the synthetic exposure mode. Another issue not specific to synthetic acquisition is the paramount importance of homogeneous illumination for speckle imaging, as inhomogeneous illumination leads to biased flow maps due to inhomogeneous shot noise contributions within the field of view[16,17]. In the synthetic exposure mode, the non-uniformity of field illumination should be minimized or corrected as differences add with the sum of the frames.

### 3.3. Relative contribution of noises as a function of the experimental conditions

An important finding from this study is that the relative contribution of noise to the speckle contrast depends on several intermingled experimental parameters. As mentioned above, the amount of available light (laser power per $mm^2$ on the sample) is essential, as detected signals far below camera saturation increase the relative contribution of the shot noise to the speckle contrast. Furthermore, the flow that is imaged is an important parameter. Slow flows observed in small arterioles capillaries and the parenchyma result in slow decorrelation of the speckles and high speckle contrast values, thus minimizing the relative contribution of the noise-related contrasts. On the contrary, imaging fast flows observed in arteries or large arterioles is much less favorable as rapid speckle decorrelation leads to low speckle contrast values. In addition, speckle contrast imaging of fast flows requires short exposure times, which could lead to sub optimal illumination. To avoid such discrepancies, we advise a systematic correction of the shot noise related contrast from all speckle contrast data.

### 3.4 Temporal resolution of synthetic exposure mode

The temporal resolution in modulated laser mode depends on the number of exposure times used and in the duration of the longest exposure time that sets the constant exposure of the camera. In a standard MESI implementation, 6 exposure times are used ranging from 1 ms up to 20 ms[18,19]. The total duration of a dataset required for the evaluation of one decorrelation time and the corresponding blood flow index is 120 ms (6x 20 ms). In the synthetic exposure mode, the temporal resolution is limited by the maximum frame rate of the camera, which is 46 fps in our setup, as a minimal 22 ms readout time is required to acquired the data from the pixels of the full sensor. Therefore, it takes almost 0.5 s to capture 20 frames at 1 ms exposure time to obtain synthetic exposure. This is a coarse temporal resolution, yet within the order of magnitude of the characteristic time of physiological blood flow changes. For instance, local increase of cerebral blood flow occurs typically within 0.5 - 1 s post sensory stimulation[20–22]. A simple way to increase the temporal resolution of the synthetic exposure mode would be to acquire data over a reduced number of pixels. For example, reducing the number of pixels read results in much higher frames rates, up to 700 fps for a region of interest of 50 x 50 pixels, leading to a temporal resolution for the blood flow index of less than 30 ms.

### 3.5 Benefits of synthetic exposure

The synthetic exposure scheme benefits form a straightforward instrumental implementation. An essential feature, regarding the camera choice is a low, stable and uniform noise. Also, an optimal, homogeneous and stable illumination is requested to ensure that the full dynamic of the camera is used and to reduce the contribution of shot noise even under unfavorable conditions (short exposure times, fast flows). The temporal resolution of the multiple exposure data is limited by the frame rate of the camera but can be adjusted easily by selecting an ROI on the sensor. As a simple and accurate implementation of the multiple exposure speckle contrast analysis, synthetic exposure is likely to pave the way for easier translation of MESI to clinical applications.

### 4. Methods

### 4.1 Speckle contrast calculation and relative flow evaluation

For all conditions, contrast images were computed using the following equation:

$$K = \frac{\sigma}{<I>} \quad (1)$$

Where $\sigma$ is the standard deviation of the intensities in the acquired image, and $<I>$ is the average intensity in the same image. A sliding spatial window of 5x5 pixels is used for the calculation of the local contrast over the whole image. The assumption behind the speckle contrast imaging of biological flows is that the flow is inversely proportional to the decorrelation time $\tau_c$ of the scatterers. Using multiple exposure contrast images, it is possible to derive $\tau_c$ while considering the contribution of static scatterers[2]. Multiple exposure speckle contrast data as a function of exposure are fitted to the following equation:

$$K(T, \tau c) = \left\{ \beta \rho^2 \ \frac{e^{-2x} - 1 + 2x}{2x^2} + 4\beta\rho(1-\rho) \ \frac{e^{-x} - 1 + x}{x^2} + v_{noise} \right\}^{½} \quad (2)$$

Where $T$ is the camera exposure time, $\beta$ is a unitless constant that accounts for spatial averaging of the spekle grains and instrumental parameters, $\rho$ is defined as the ratio of the intensity contribution of mobile scatterers over the total intensity due to mobile and static scatterers, x is the ratio of T over $\tau_c$, and $v_{noise}$ accounts for experimental noises contribution.

The correlation time $\tau_c$ is derived from the fit and the relative blood flow index (BFI) is defined as

$$BFI = \frac{1}{\tau_c} \quad (3)$$

### 4.2 Acquisition modes and imaging set up

The imaging set-up is showed in figure 6. A 634 nm laser diode with a maximum output power of 419 mW is used. Images were acquired through a stereomicroscope (Leica MZ16) with a genI sCMOS camera (Orca Flash 2.8, Hamamatsu, Japan). It has a 42% spectral sensitivity at 634 nm, a typical read noise of 3e- and a dark noise of 1e-. The microfluidic channels fabrication and flow control set-up were described in details previously[6] Images were acquired for various flows in microfluidic channels with a rectangular section of 300 μm widths and 75 μm of height. We used 2% intralipid as the flowing media since its optical scattering properties are close to those of blood. Flows from 1 μl.min-1 up to 7 μl.min-1 are controlled using a pressure controller with inline flow feedback (Fluigent, France).

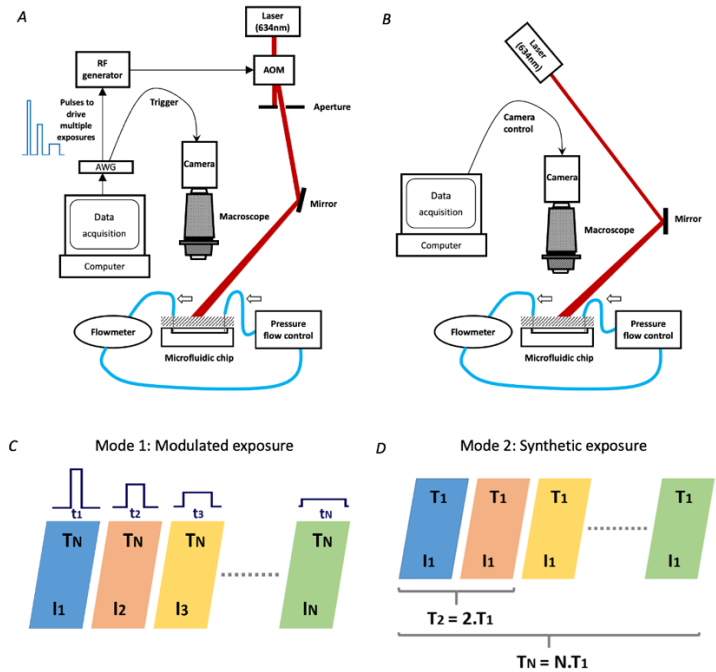

Figure 6. Imaging set up and acquisition modes A) Modulated laser mode set-up (mode 1). AWG = arbitrary waveform generator, AOM = acousto-optic modulator B) Synthetic exposure mode set-up (mode 2) C) Modulated laser acquisition scheme. N is the number of different exposures at which the speckle patterns are recorded. The camera exposure time T is fixed and the laser is modulated in duration t and intensity I  D) Synthetic exposure acquisition scheme.  The laser is operated in continuous mode, its intensity I is fixed, the camera exposure time is fixed and set to the shortest exposure time $T_1$.

### 4.3 Multiple exposure acquisition modes

Multiple exposure speckle data were obtained for the following exposure times,1, 2, 5, 7, 10, 20 ms. In a first approach, called laser modulation method (Figure 6C, mode 1), the acquisition of multiple exposure relies on a camera exposure fixed at 50 ms synchronized to laser pulses varying in duration and intensity using an acousto-optic modulator. The width of each pulse is set to the desired exposure duration, while its amplitude is set so that the pulses energies are equalized. The shortest exposure time requires the maximum amplitude, whereas the longest exposure is associated with the lowest amplitude. Contrast images are then calculated for all exposure times using equation (1) and $\tau_c$ is derived from a fit of the contrast data to equation (2). In a second approach, called synthetic exposure (Figure 6D, mode 2), contrast images are generated using synthetic exposures as follows: the camera exposure time is set to 1 ms and the laser diode power is adjusted to ensure that the maximum dynamic of the sensor is used. 50 independent frames at 1 ms are acquired and contrast images for $T$: 1, 2, 5, 7, 10, 20 are computed as the sum of the corresponding number of 1 ms-frames. The interframe time is fixed to 22 ms by the readout speed of the camera electronics consequently uncorrelated short exposure frames (1 ms) are summed to create longer exposure times.

### 4.4 Noise evaluations and correction

Speckle contrast relies on the statistical properties of the local signal intensity and is calculated using equation (1) and a 5x5 spatial kernel[6]. However, there are several noise contributions arising from the detector including dark noise, readout noise and shot noise. Here we evaluate these contributions to obtain the noise-corrected speckle contrast $K_{corr}$. First, images are acquired with the laser source off to account for dark and readout noise simultaneously. From these images, the speckle contrast $K_{dark}$ corresponding to the dark and readout fluctuations is derived following Equation (4).

$$K_{dark} = \frac{\sigma_{dark}}{\langle I_{dark} \rangle} \tag{4}$$

Second, the mean of a stack of 50 dark images is subtracted pixelwise from the raw speckle image to correct for the noise contribution to the signal intensity prior to contrast calculation following Equation (1).

$$I_c = I_{raw} - \langle I_{dark} \rangle \tag{5}$$

However, this operation does not correct for the added variability, as subtraction of one image to another one results in adding the variance of the intensities for each pixel. Thus, the contribution of dark noise fluctuation $\sigma_d^2$ to the variance should be considered.

The contribution of the shot noise obeys to Poisson's statistics with a variance $\sigma_s^2 = \langle I_c \rangle$ equal to the mean resulting in a shot noise related contrast contribution calculated as:

$$K_{shot} = \frac{1}{\sqrt{I_c}} \tag{6}$$

Finally, the speckle contrast corrected for both the shot noise and dark-readout noise is given as[14]

$$K = \sqrt{\frac{\sigma_c^2 - \sigma_s^2 - \sigma_d^2}{\langle I_c \rangle^2}} \tag{7}$$

**Authors contributions statements**
FP and MC designed the study and developed the synthetic exposure set-up. MC conducted the imaging experiments. MC and FP analyzed the data. MC and FP wrote and edited the manuscript. All the authors have reviewed the entire manuscript.

**Competing interests**
The author(s) declare no competing interests.

**Data availability**
The datasets generated during the current study are available from the corresponding author on reasonable request

# Figures captions

**Figure 1** Camera dark and readout noise homogeneity and stability A) Representative dark image B) Corresponding histogram of grey levels. Insert is a zoom on non-zero histogram levels C) Spatial uniformity of noise, the sensor was divided in 16 regions of identical dimensions numbered from 1 to 16 D) Temporal stability of noise evaluated as the mean grey level of the whole sensor.

**Figure 2** A) Raw, dark and readout, and shot noise standard deviations in the modulated laser mode as a function of exposure time. B) Same as A) for the synthetic exposure mode C) Raw, dark and readout and shot noise speckle contrast for the modulated laser mode D) Same as C) for the synthetic exposure mode.

**Figure 3** Effect of illumination level and exposure time on the shot noise contribution to the speckle contrast in synthetic exposition mode A) Effect of the detected light intensity; empty dots (raw contrast) and filled dots (noise corrected contrast) correspond to an exposure time of 1 ms ; empty triangles (raw contrast) and filled triangles (noise corrected contrast) correspond to an exposure time of 25 ms B) Effect of the exposure time; empty squares (raw contrast) and filled squares (noise corrected contrast) correspond to an illumination of 2 mW/cm$^2$; empty diamonds (raw contrast) and filled diamonds (noise corrected contrast) correspond to an illumination of 223mW/cm$^2$ . Insert : Relative error due to the shot noise speckle contrast as a function of exposure time for different illumination levels; squares, dots and diamonds correspond respectively to illuminations at 20, 100 and 223mW/cm$^2$.

**Figure 4** Relative error on speckle contrast due to noises contributions as a function of exposure time for different flows of intralipid 2% in a 300 µm diameter channel. Data have been acquired in synthetic exposure, with frames of 1 ms exposure time, and under optimal illumination using the full dynamic of the camera sensor.

**Figure 5**. Representative comparison between mode 1 (Laser Modulated Method) and mode 2 (Synthetic Exposure) **A)** Speckle contrast images of intralipid 2% flowing at 3µl/min in a 300 µm diameter channel for 6 exposure times (mode 1) **B)** Same as (A) for mode 2 **C)** Speckle contrast versus exposure time **D)** normalized correlation times for mode 1 as a function of those of mode 2.

**Figure 6**. Imaging set up and acquisition modes A) Modulated laser mode set-up (mode 1). AWG = arbitrary waveform generator, AOM = acousto-optic modulator B) Synthetic exposure mode set-up (mode 2) C) Modulated laser acquisition scheme. N is the number of different exposures at which the speckle patterns are recorded. The camera exposure time T is fixed and the laser is modulated in duration t and intensity I D) Synthetic exposure acquisition scheme. The laser is operated in continuous mode, its intensity I is fixed, the camera exposure time is fixed and set to the shortest exposure time $T_1$.